\documentclass[reprint,twocolumn,amsmath,amssymb,aps,pre,showpacs]{revtex4-1}
\usepackage[english]{babel}
\usepackage{graphicx}
\usepackage{bm}

\newcommand{\bk}{\mathbf{k}}
\newcommand{\bn}{\mathbf{n}}
\newcommand{\bq}{\mathbf{q}}
\newcommand{\bQ}{\mathbf{Q}}
\newcommand{\bB}{\mathbf{B}}
\newcommand{\id}{\mathrm{d}}
\newcommand{\br}{\mathbf{r}}
\newcommand{\bR}{\mathbf{R}}
\newcommand{\bG}{\mathbf{G}}
\newcommand{\bg}{\mathbf{g}}
\newcommand{\bv}{\mathbf{v}}
\newcommand{\bx}{\mathbf{x}}

\newcommand{\bff}{\bm{f}}
\newcommand{\bu}{\mathbf{u}}
\newcommand{\bOmega}{\bm{\Omega}}

\begin{document}
\title{Attraction/repulsion switching of non-equilibrium depletion interaction\\
caused by blockade effect in gas of interacting particles. II}

\author{O.V.~Kliushnychenko}
\affiliation{Institute of Physics, NAS of Ukraine, Prospect Nauky 46, 03028 Kiev, Ukraine}

\author{S.P.~Lukyanets} \email[Email address: ]{lukyan@iop.kiev.ua}

\affiliation{Institute of Physics, NAS of Ukraine, Prospect Nauky 46, 03028 Kiev, Ukraine}

\begin{abstract}
The effect of concentration-dependent switching of the non-equilibrium depletion interaction between obstacles in a gas flow of interacting Brownian particles is presented. When increasing bath fraction exceeds half-filling, the wake-mediated interaction between obstacles switches from effective attraction to repulsion or vice-versa, depending on the mutual alignment of obstacles with respect to the gas flow. It is shown that for an ensemble of small and widely separated obstacles the dissipative interaction takes the form of induced dipole-dipole interaction governed by an anisotropic screened Coulomb-like potential. This allows one to give a qualitative picture of the interaction between obstacles and explain switching effect as a result of changes of anisotropy direction. The non-linear blockade effect is shown to be essential near closely located obstacles, that manifests itself in additional screening of the gas flow and generation of a pronounced step-like profile of gas density distribution. It is established that behavior of the magnitude of dissipative effective interaction is, generally, non-monotonic in relation to both the bath fraction and the external driving field. It has characteristic peaks corresponding to the situation when the common density ``coat'' formed around the obstacles is most pronounced. The possibility of the dissipative pairing effect is briefly discussed. All the results are obtained within the classical lattice-gas model.
\end{abstract}



\maketitle

\section{\label{sec:intro} Introduction}

Motion of inclusions or probe-particles through a medium is accompanied by the medium perturbation (e.g., perturbation of its density) that can manifest itself in the form of wakes. The medium perturbation can, in turn, induce a non-equilibrium interaction between the inclusions. Such interaction is responsible, in particular, for the coherent part of the collective friction force as well as for possible formation of dissipative structures in an ensemble of the inclusions. The nature of the medium perturbation and the properties of the induced non-equilibrium interaction are defined by the properties of the medium (e.g., by its nonlinearity) and the mechanism of energy losses. The perturbation can lead to generation of vortices, Cherenkov radiation, or local phase transitions (some more effect can be found in hydrodynamics \cite{kelvin_lxxi._1905,lamb_hydrodynamics_1975,birkhoff_jets_1957,khair_motion_2007,sriram_out--equilibrium_2012,sriram_two_spheres_2015}, optics \cite{couairon_femtosecond_2007}, plasma physics \cite{landau_electrodynamics_1984,ritchie_wake_1982,ritchie_wake_1976,morfill_complex_2009,tsytovich_nonlinear_2013,tsytovich_self_2015}, quantum liquids and Bose condensates \cite{pines_theory_1966,gladush_generation_2007,kamchatnov_stabilization_2008,mironov_structure_2010,roberts_casimir-like_2005,lychkovskiy_2015,carusotto_2013}).

In dissipative media, the induced non-equilibrium interaction between inclusions can conditionally be divided into reactive and dissipative parts. In the simplest case, when the speed of a probe-particle is rather small (e.g., smaller than the speed of sound in a medium and the hydrodynamic effects can be neglected, the medium perturbation can be described in the diffusive approximation \cite{forster_1975}. The diffusive wake may be of large spatial and temporal extensions with power-law damping (see \cite{benichou_stokes_2000,benichou_force_2001,benichou_biased_2013,benichou_2015,demery_drag_2010}), which is an evidence of long-time memory of the medium about the particle passage. The long-living wakes of individual particles lead, in turn, to a long-range effective dissipative interaction between the particles \cite{dzubiella_depletion_2003}. This can be qualitatively described using the linear response approximation \cite{pines_theory_1966,linear_response}. However, the linear response approximation, giving a qualitative picture of medium perturbation, leads to incorrect results for wakes, dissipative interaction and, in general, does not give adequate description of non-linear media \cite{comment}.

In the present paper, we will be interested in the dissipative interaction between inclusions induced by their wakes in a nonlinear medium, resorting to an example of a Brownian gas with short-range inter-particle repulsion (the hard-core interaction). In this case, the dissipative interaction between inclusions is often called the non-equilibrium depletion or entropic interaction (e.g., \cite{dzubiella_depletion_2003,Sasa_2006,khair_motion_2007}).
At equilibrium, the depletion interaction is usually short-range; its spatial range is of the order of the characteristic length scale of the medium particles \cite{lekkerkerker_2011,crocker_entropic_1999}. In contrast, the non-equilibrium forces between impurities may exhibit long-range behavior due to a long-living diffusive wake induced by their motion \cite{demery_drag_2010,demery_2014,benichou_stokes_2000,benichou_force_2001,benichou_biased_2013,dzubiella_depletion_2003}.
In addition, such forces often have unusual properties, e.g., they violate the Newton's third law \cite{dzubiella_depletion_2003,Sasa_2006,pinheiro_2011,Ivlev_2015}. The non-Newtonian behavior of the non-equilibrium depletion force was demonstrated at low gas concentrations \cite{dzubiella_depletion_2003}, when interaction between gas particles is negligible.

To describe the non-equilibrium depletion force for a gas of interacting particles, we turn to the simplest model of a lattice gas, when each lattice site can be occupied by only one particle. Even such a short-range interaction results in a number of unexpected kinetic effects, e.g., the ``back correlations'' effect \cite{tahir-kheli_correlated_1983}, drifting spatial structures \cite{schmittmann_statistical_1995,leung_drifting_1997,hipolito_effects_2003}, effects of ``nega\-tive'' mass transport \cite{Lukyanets2010,argyrakis_negative_2009,efros_negative_2008}, induced long-time correlations \cite{kliushnychenko_induced_2013}, and the dissipative pairing effect for tracers passing through a lattice gas \cite{mejia-monasterio_bias-_2011}. Increasing of gas concentration (bath fraction) leads to enhancement of the role of interaction between gas particles. As was shown in \cite{kliushnychenko_blockade_2014}, this implicates significant changes in the shape of wake of a fixed inclusion (or an obstacle) in a gas flow --- wake inversion. In turn, the wake-mediated interaction between obstacles should be sensitive to the crucial transformation of the wake structure.

In this paper we examine how the short-range repulsive interaction of gas particles affects the behavior of dissipative forces between obstacles embedded into the gas flow. In particular, we show that increasing of gas concentration can lead to the switching, or sign change, of the effective dissipative interaction between obstacles to its opposite, e.g., from attraction to repulsion or visa-versa. This effect is entailed by the obstacle wake inversion considered in \cite{kliushnychenko_blockade_2014}. For closely located obstacles, the interaction of gas particles is shown to provoke non-linear blockade effect, that results in formation of a common ``coat'' of gas density perturbation around the obstacles with a pronounced step-like behavior of its distribution. In turn, the common non-linear coat can signify the dissipative pairing of the obstacles, see \cite{mejia-monasterio_bias-_2011}. In the case of small and widely separated inclusions, when the non-linear effects are less significant, we show that the dissipative interaction between them belongs to the type of induced dipole-dipole (generally multipole) interaction associated with anisotropic screened Coulomb-like potential. To demonstrate the above mentioned phenomena, we use the mean-field and the long-wavelength approximations, neglecting the short-range correlations and fluctuations in the gas, see \cite{kliushnychenko_blockade_2014,Lukyanets2010}.

Our paper is organized as follows: In Sec.~\ref{sec:model} we specify the kinetic equations to be used and briefly discuss the employed approximations. The main results on dissipative forces are contained in Sec.~\ref{sec:inversion}. In Sec.~\ref{sec:asymptotics}, the case of small (point-like) inclusions is considered in the linear flow approximation. In Sec.~\ref{sec:nl}, the non-linear blockade effect (i.e., screening of gas flow) is discussed for large and closely located obstacles. In Sec.~\ref{sec:dfswitch}, the case of two moderately separated obstacles is considered numerically for two spatial configurations. Sec.~\ref{sec:conclusion} briefly summarizes obtained results. Appendixes contain the outlines of two analytic approaches used in Sec.~\ref{sec:asymptotics}: a na\"{\i}ve one (Appendix~\ref{A1}), giving a rough sketch of the dissipative interaction behavior, and a more sophisticated one (Appendix~\ref{A2}), based on the single-layer potential method for inclusions with sharp boundaries.

\section{\label{sec:model} Model}

As was shown in \cite{kliushnychenko_blockade_2014,Lukyanets2010}, an obstacle in a lattice gas flow can be considered as a limiting case of a two-component gas: one of the components is static while the other one is mobile and driven by a uniform external field. We employ the simplest model of a two-component lattice gas, when each lattice site can be occupied by only one particle, see \cite{tahir-kheli_correlated_1983}. Kinetics of a multicomponent lattice gas is defined by the jumps of its particles to the neighboring vacant sites. The variation of the $i$th site occupancy by the particles of the $\alpha$th sort during the time interval $\Delta t$, $\tau_0\ll\Delta t\ll\tau_l$ ($\tau_0$ is the duration of a particle jump to a neighboring site and $\tau_l$ being the lifetime of a particle on a site), is described by the standard continuity equation (see, e.g., \cite{chumak_diffusion_1980,tahir-kheli_correlated_1983})
\begin{equation}\label{balance+}
n_i^\alpha (t+\Delta t)-n_i^\alpha(t)=\sum_j
\left(J^\alpha_{ji}-J^\alpha_{ij}\right) + \delta J_i^\alpha,
\end{equation}
where $\alpha$ and $\beta$ label the particle species and $n_i^\alpha=0,1$ are the local occupation numbers of the $\alpha$th particles at the $i$th site. $J^\alpha_{ij}=\nu^\alpha_{ij}n_i^\alpha \left(1-\sum_\beta n_j^\beta\right) \Delta t$ gives the average number of jumps of the $\alpha$th particles from site $i$ to a neighboring site $j$ per time $\Delta t$. $\nu_{ij}^\alpha=\nu^\alpha$ is the mean frequency of these jumps. The term $\delta J_i^\alpha=\sum_j(\delta J_{ji}^\alpha-\delta J_{ij}^\alpha)$ stands for the Langevin source that is defined by the fluctuations $\delta J_{ji}^\alpha$ of the number of jumps between sites $j$ and $i$ during $\Delta t$ \cite{chumak_diffusion_1980}.
These fluctuations are caused by fast, compared to the time scale $\Delta t$, processes and will be neglected for simplicity. It means that we disregard the fluctuation-induced forces.

In what follows we consider only two components, mobile and static, which are labeled by $n$ and $u$, respectively. In the absence of external fields we suggest for a regular lattice that $\nu_{ji}^n=\nu=\mathrm{const}$ for the component $n$, while the component $u$ is assumed to be at rest, $\nu_{ji}^u=0$. The presence of a driving field leads to asymmetry of the particle jumps. Assuming the activation mechanism of the jumps and a weak driving field $\bG$, frequency $\nu_{ji}$ may be written as $\nu_{ji}^n\approx\nu[1+(\bG,\br_i-\br_j)/(2kT)]$, or
$\nu^\pm\approx\nu\pm\delta\nu$, where $\nu^+$ and $\nu^-$ denote the jump frequencies along and against the field, respectively. $\delta\nu=\nu \ell|\bG|/(2kT)$ ($\ell$ is the lattice constant), condition $\ell|\bG|/(2kT)<1$ is assumed to be satisfied.

Equations for the average local occupation numbers can be obtained from Eqs.~(\ref{balance+}) using the local equilibrium approximation (the Zubarev approach) \cite{chumak_diffusion_1980,zubarev_nonequilibrium_1974} which coincides, in our case,  with the mean-field approximation \cite{leung_novel_1994}. Introducing time derivatives \cite{richards_theory_1977}, in the long-wavelength approximation (see \cite{schmittmann_statistical_1995,leung_drifting_1997,hipolito_effects_2003,Lukyanets2010}) the macroscopic kinetics of the mobile component $n$ is given by the equation
\begin{equation}\label{eq:lw}
  \partial_\tau n = \nabla^2n-\nabla(u\nabla n - n\nabla u)-(\bg,\nabla)[n(1-u-n)],
\end{equation}
where $n=n(\br,\tau)$ and $u=u(\br)$ are the average occupation numbers of the two components at the point $\br$ ($0\leq n\leq1$ and $0\leq u\leq1$) and $\bg=\ell\bG/(2kT)$. Here, we have introduced the dimensionless spatial coordinate $\br/\ell\rightarrow\br$ and time $\tau=\nu t$, and $\partial_\tau$ stands for the partial time derivative. Note that equations of the form (\ref{eq:lw}), as well as their generalizations for two- and multicomponent systems, also appear in the problems of nonlinear cross-diffusion with size exclusion \cite{burger_2010}, diffusion in monolayers of reagents on the surface of a catalyst \cite{gorban_2011} and serve as a model of fast ionic conductors \cite{schmittmann_statistical_1995}.

In the non-equilibrium case, there are various approaches to introduce the dissipative force (or interaction) between inclusions via the Brownian gas environment. The approaches are not equivalent to each other and may lead to different results in general, see \cite{Sasa_2006}. To introduce the force acting on an obstacle, we first consider a point-like inclusion (impurity) occupying a lattice site $\bR_j$ with a given interaction potential $U(\br_i-\bR_j)$ between the inclusion and a particle of the lattice gas at site $\br_i$. Then, Hamiltonian of the lattice gas in the presence of impurities is written as $H=H_0+H_{\mathit{int}}$, where $H_0$ is the Hamiltonian of the lattice gas without inclusions, and $H_{\mathit{int}}=\sum_{ij}n_iU(\br_i-\bR_j)$ describes interaction between gas particles and impurities, $n_i=0,1$ is the occupation number of site $\br_i$.

At equilibrium, the total force acting on the $j$th inclusion can be written as (see \cite{Sasa_2006})
\begin{eqnarray}\label{eq:eforce1}
  \bff_j^{\mathit{eq}}&=&\left\langle -\frac{\delta}{\delta\bR_j}H_{\mathit{int}}\right\rangle
  = \sum_{\{n\}} \left(-\frac{\delta}{\delta\bR_j}H_{\mathit{int}}\right)\rho(\{n\}) \\
  &=& \sum_i\langle n_i\rangle\frac{\delta}{\delta\br_i}U(\br_i-\bR_j),\label{eq:eforce2}
\end{eqnarray}
where $\rho(\{n\})$ is the equilibrium probability (or statistical operator in the matrix representation \cite{chumak_diffusion_1980}) of a given occupancy configuration $\{n\}$
\begin{equation}\label{distrfunc}
  \rho(\{n\},0)=Z^{-1}\exp(-H\{n\}/kT)
\end{equation}
and $Z=\sum_{\{n\}}\exp(-H\{n\}/kT)$; $\langle n_i\rangle$ is the mean occupation number at site $\br_i$ that describes the equilibrium distribution of gas concentration. The force $\bff_j^{\mathit{eq}}$, Eqs.~(\ref{eq:eforce1}) or (\ref{eq:eforce2}), can be expressed in terms of the gas free energy $F=-kT\ln Z$ as
\begin{equation}\label{eq:neforce1}
  \bff_j^{\mathit{eq}}=-\frac{\delta}{\delta\bR_j}F.
\end{equation}
This relation is often used to define the equilibrium depletion force \cite{Asakura_1954,Asakura_1985}.

In this paper, we use another approach based on expression (\ref{eq:eforce1}) written with non-equilibrium statistical operator $\rho_t(\{n\})$ (see \cite{Sasa_2006})
\begin{equation}\label{eq:neforce3}
  \bff_j^{\mathit{neq}}=\sum_{\{n\}}\left( -\frac{\delta H_{\mathrm{int}}}{\delta\bR_j}\right)\rho_t(\{n\})
  = \sum_i \langle n_i\rangle_t\frac{\delta}{\delta\br_i}U(\br_i-\bR_j),
\end{equation}
where $\rho_t(\{n\})$ obeys a master equation for the hopping process, see \cite{bitbol_forces_2011}, and $\langle n_i\rangle_t=\sum_{\{n\}} n_i\rho_t(\{n\})$ is non-equilibrium gas concentration. Yet another approach consists in generalizing Eq.~(\ref{eq:neforce1}) to the non-equilibrium case by introducing an effective non-equilibrium potential or non-equilibrium free energy for a gas \cite{schmittmann_statistical_1995,bitbol_forces_2011,likos_effective_2001,gouyet_descr_2003,Sasa_2006}. As was shown in \cite{Sasa_2006}, these two definitions of the non-equilibrium force are not equivalent. Representation (\ref{eq:neforce3}) for the force exerted by gas particles on an obstacle is similar to the hydrodynamic definition of the force which, in particular, was used in \cite{dzubiella_depletion_2003} to describe the non-equilibrium depletion interaction between obstacles in a gas of non-interacting particles. Here, we use  representation (\ref{eq:neforce3}) to describe the non-equilibrium depletion forces acting between obstacles via gas perturbation.

In the continuum limit and the mean-field approximation, $\bff_j^{\mathit{neq}}$ takes the form
\begin{equation}\label{eq:new}
  \bff_j^{\mathit{neq}} = -\int U(\br-\bR_j)\nabla_\br n(\br,t)\,\id\br,
\end{equation}
where $n(\br,t)=\langle n(\br)\rangle_t$.
When the obstacle is a cluster formed by particles of the second (heavy) gas component, potential $U(\br)$ describes the concentration distribution of that component and $n(\br,t)$ obeys Eq.~(\ref{eq:lw}) obtained in the long-wavelength approximation. In what follows, to separate out the contribution of the gas perturbation $\delta n(\br,t)$ induced by the gas flow (or the external field $\bg$) from the total force (\ref{eq:new}), we consider the quantity
\begin{equation}\label{eq:new1}
  \bff_j = -\int U(\br-\bR_j)\nabla_\br \delta n(\br,t)\,\id\br,
\end{equation}
where $\delta n(\br,t) =n(\br,t)-n_0 (\br)$, $n_0(\br)$ is the equilibrium concentration distribution, and $n_0(\br\rightarrow\infty)\rightarrow n_0\equiv\mathrm{const}$ stands for the average equilibrium concentration of gas (fraction of the full lattice occupation, $0\leq n_0\leq1$).

In the case of inclusion with a sharp boundary, $\bff_j$ takes the conventional form
\begin{equation}\label{eq:force}
  \bff_j=-\int_{S_j}\bn(\br)\delta n(\br)\,\id \br,
\end{equation}
where $S_j$ is the surface of $j$th inclusion and $\bn(\br)$ is its exterior normal at the point $\br$. In what follows, we will be interested in non-equilibrium steady-state interaction, i.e., in the limiting case $t\rightarrow\infty$. We will use the lattice gas model (\ref{balance+}) in the mean-field approximation (neglecting the fluctuation part) and its continuum version (\ref{eq:lw}) to describe the character of the dissipative interaction between obstacles.

\section{\label{sec:inversion}
Inversion of wake and switching of dissipative interaction}

\begin{figure*}
\includegraphics[width=2\columnwidth]{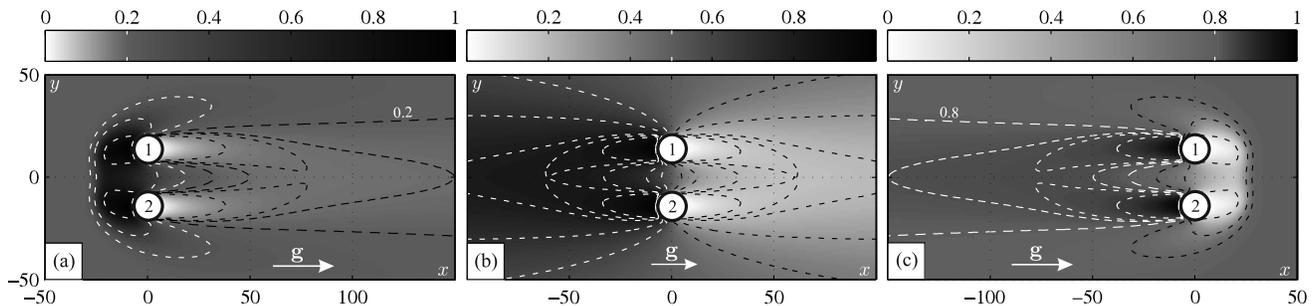}
\caption{\label{bconf} \textit{Transverse alignment.}
Steady-state concentration distributions (average occupation numbers) of the gas particles $n(x,y)$ near the obstacles, evaluated numerically within the mean-field approximation of Eq.~(\ref{balance+}) in the 2D case, correspond to various equilibrium concentrations $n_0 = 0.2$ (a), $0.5$ (b), and $0.8$ (c). Panels (a) to (c) illustrate three different regimes of the dissipative interaction: (a) effective repulsion ($f^y_{12}>0$, $f^y_{21}<0$, $|f^y_{21}|=|f^y_{12}|$), (b) no interaction ($f^y_{21}=f^y_{12}=0$), (c) effective attraction ($f^y_{21}>f^y_{12}$). The external field $\bg$ ($|\bg|=0.5$) is directed along the $x$-axis; the impermeable ($\bar u=1$) circular obstacles are of radius $a=7$ (in units of $\ell$); the distance between the obstacle centers equals $4a$. The gray background corresponds to the equilibrium gas concentration $n_0$ for every contour plot, in consistence with the color bars. Spatial coordinates are in units of $\ell$.}
\end{figure*}

In this section we consider how the gas particle interaction and non-linear screening of gas flow affect the behavior of dissipative forces acting on obstacles. In particular, we show that the short-range repulsive interaction between gas particles can lead to switching of non-equilibrium depletion interaction between obstacles, e.g., from effective repulsion to attraction, as the equilibrium gas concentration $n_0$ increases. Such interaction switching is directly related to the obstacle wake inversion effect considered in \cite{kliushnychenko_blockade_2014}. As was shown in \cite{kliushnychenko_blockade_2014}, increasing of gas concentration $n_0$ leads to drastic transformation of the inclusion wake structure: typical wake \cite{dzubiella_depletion_2003,khair_motion_2007,sriram_out--equilibrium_2012,benichou_biased_2013} with an extended depleted region behind the inclusion and localized dense region in front of it [at $n_0<1/2$, see Fig.~\ref{bconf}(a)],
acquires an unusual inverted structure with an extended dense region ahead of the obstacle and a localized depleted region behind [at $n_0>1/2$, Fig.~\ref{bconf}(c)]. Note that the possibility of wake and force switching can be easily shown by using the ``hole-particle'' symmetry of Eq.~(\ref{eq:lw}), see \cite{kliushnychenko_blockade_2014,kolomeisky_asymmetric_1998}.

Switching of the wake ``direction'' at high gas concentration is caused by the enhancement of the role of interaction between gas particles, in particular, can lead to the non-linear blockade effect. This effect is significant near the obstacle surface, especially for large and for closely located ones. For a relatively large obstacle and sufficiently high concentration $n_0$, the gas flow generates a dense region ahead of the obstacle as the gas particles have no time to leave this zone via lateral diffusion. Such a strong accumulation of the gas particles locally enhances the significance of the interaction between them, so that the dense region ahead of the obstacle has to grow. Similar behavior arises for closely located obstacles when their individual density perturbation ``coats'' overlap leading to formation of a common ``coat'' around them and to additional screening of the gas flow. The latter means that peculiarities of dissipative interaction between closely located obstacles are determined by the non-linear blockade effect for which the term $\sim n^2$ in Eq.~(\ref{eq:lw}) is responsible. We consider these non-linear effects numerically on the basis of mean-field version of Eq.~(\ref{balance+}), neglecting gas fluctuations.

In the particular case of relatively small and distant obstacles, the interaction between gas particles can be taken into account in the linear approximation \cite{kliushnychenko_blockade_2014}. That approximation allows one to obtain analytical expressions for the asymptotic behavior of both the density perturbation far from obstacles and the dissipative interaction between them. Now we proceed to this case in the subsection below.

\subsection{\label{sec:asymptotics}
The asymptotic behavior of density perturbations and dissipative forces for distant inclusions}

In what follows we consider a non-equilibrium steady-state problem by setting $\partial_\tau n=0$ in Eq.~(\ref{eq:lw}):
\begin{equation}\label{eq:steady}
  \nabla^2 n - U\nabla^2 n + n\nabla^2 U - (\bg,\nabla)n(1 - n - U)=0,
\end{equation}
where obstacles are given by a distribution $U$ of the heavy gas component. Far from small obstacle (whose size is comparable with lattice constant) the density distribution $n=n_0+\delta n$ weakly deviates from the equilibrium one $n_0$ \cite{kliushnychenko_blockade_2014}. In this case, interaction between gas particles is less significant and the drift term in Eq.~(\ref{eq:lw}) can be written in the linear approximation $n^2\approx n_0^2 + 2n_0 \delta n$.

Simple analytical expressions for density perturbation and dissipative forces for the ensemble of widely separated small obstacles can be obtained using the qualitative approach described in Appendix~\ref{A1}. This approach is similar to that based on the method of molecular field that was used to describe the elastic interaction of colloidal particles in a liquid crystal, see \cite{Lev_Interaction_1999}. In particular, the gas density perturbation far from an isolated obstacle can be written as
\begin{equation}\label{eq:deltaen}
    \delta n(\br) \sim \left(\overline{\bOmega}, \nabla_{\br}\right) G(\br),
\end{equation}
where $G(\br)$ is anisotropic screened Coulomb-like potential that takes the form
\begin{equation}\label{eq:GF3D}
  G(\br) = \frac{1}{4\pi}\frac{e^{-q|\br|+\bq\br}}{|\br|}
\end{equation}
in 3D case, and
\begin{equation}\label{eq:GF2D}
  G(\br) = \frac{1}{2\pi}e^{\bq\br}K_0(q|\br|)
\end{equation}
in 2D case. Here, $K_0$ is the modified Bessel function, vector $\bq=(1/2-n_0)\bg$ determines the preferable direction of screening and depends on external sweeping field $\bg$ (or gas flow) and on equilibrium gas concentration $n_0$ (bath fraction). $\overline{\bOmega}$ plays a role of the molecular field or an average flux near obstacle, see Appendix~\ref{A1}.

At low concentrations of gas ($n_0<1/2$), the dense region of the gas ahead of the inclusion is described by an exponential asymptotics, while the asymptotics of the depletion region behind the inclusion is power-law. When gas concentration increases and $n_0$ becomes greater than $1/2$, the anisotropy vector $\bq=(1/2-n_0)\bg$ changes its direction to the opposite. It means that switching of the wake direction occurs together with corresponding switching between the exponential and power-law asymptotics. The distribution $\delta n(\br)$, related to the anisotropic screened Coulomb-like potential, formally describes a ``medium polarization'' around the inclusion induced by an asymmetrical ``dipole'' (see, e.g., Fig.~\ref{bconf}).

In the case of ensemble of distant inclusions, the force exerted by the $i$th inclusion on the $k$th one can be roughly estimated as (see Appendix~\ref{A1})
\begin{equation}\label{eq:dforcekj}
\bff_{ki} \sim  -\nabla_{\bR_k}\left(\overline\bOmega_i, \nabla_{\bR_k}\right) G(\bR_k-\bR_i),
\end{equation}
where $\bR_k$ is the center of the $k$th inclusion. The anisotropic screened Coulomb potential $G$, giving the asymmetrical form of the obstacle wake (\ref{eq:deltaen}), naturally leads to the non-Newtonian character of dissipative forces acting between obstacles, $\bff_{ki}\neq -\bff_{ik}$. As seen from Eqs.~(\ref{eq:deltaen}) and (\ref{eq:dforcekj}), the asymptotic behaviors of the density  perturbation and  the dissipative forces acting between widely spaced small inclusions are defined by the moments of a screened anisotropic Coulomb potential. The local density perturbation around an obstacle is formed by an effective flow $\overline \bOmega$  (molecular field) that is determined by the external flow and flows induced by gas density perturbations of all the inclusions. The latter means that the interaction between two inclusions cannot be separated out of the influence of all other inclusions. This is a general property of a nonlinear response or nonlinear systems, see \cite{Lev_Interaction_1999}. The employed approach allows one to consider the non-linear response and to represent expressions for $\delta n$ and $\bff_{kj}$ in the form similar to that given by the linear response for moving probe-particles (cf. note \cite{linear_response}), the only difference is that asymptotic behaviors are associated with anisotropic screened Coulomb potential instead of Coulomb one, $|\br|^{-1}$, and with mean gas flow (mean field) near inclusion instead of velocity of a probe-particle. However, Eqs.~(\ref{eq:deltaen}) and (\ref{eq:dforcekj}) are obtained within somewhat na\"{\i}ve approach and give only a qualitative picture of dissipative interaction.

More rigorous results for the asymptotics behavior can be obtained within the single-layer potential approach for inclusions with sharp boundaries. Representation of solution for $\delta n$ in the form of single-layer potential was proposed in \cite{kliushnychenko_blockade_2014} to describe the gas density perturbation around single obstacle in 2D case. In this paper, we use this representation and its multipole expansion (Appendix~\ref{A2}) to find a general form of asymptotic behavior of dissipative forces for widely separated obstacles.
Particularly, in 3D case this method gives the following asymptotic behaviors:
\begin{equation}\label{n13D}
    \delta n(\br)\approx \frac{e^{-q|\br|+\bq\br}}{|\br|}\tilde I(\br,\bq)
\end{equation}
for density perturbation caused by a small isolated obstacle and
\begin{equation}\label{eq:f2}
  \bff_{ki}\approx -\frac{e^{-q|\br_{ki}|+\bq\br_{ki}}}{4\pi|\br_{ki}|}
  \int_{S_k}\bn(\bx_k) I\left(\br_{ki},\bq,\bx_k\right)
  \,\id\bx_k
\end{equation}
for the dissipative force exerted by $i$th inclusion on $k$th one in the dipole approximation, when the distance $|\br_{ik}|=|\bR_i - \bR_k|$ between inclusions is much larger than their radii $a_i (a_k)\sim \ell$, $\bn(\bx_k)$ is the exterior normal at the point $\bx_k$ on the surface of the $k$th inclusion. For simplicity, we have considered spherical obstacles, $S_k$ is the surface of the $k$th inclusion ($|\bx_k|=a_k$).
Functions $\tilde I(\br,\bq)$ and $I\left(\br_{ki},\bq,\bx_k\right)$  have a power-law dependence on $1/r$ and $1/r_{ki}$, respectively [see Eqs.~(\ref{eq:A2:Is}) and (\ref{eq:A2:Iki})]. In particular, function $I\left(\br_{ki},\bq,\bx_k\right)$ can be represented in general form as
\begin{equation}
  I=A\left(\bq,a_i,\bx_k,\right)+\bB\left(\bq,a_i,\bx_k,\right)\left(\bq-q\frac{\br_{ki}}{|\br_{ki}|} -\frac{\br_{ki}}{|\br_{ki}|^2}\right),
\end{equation}
where $A$ and $\bB$ are determined only by the obstacle surface and external field $\bg$.

In 2D case, $\delta n$ and $\bff_{ki}$ are determined by the potential $\exp(\bq\br)K_0(qr)$, Eq.~(\ref{eq:GF2D}), having the asymptotic behavior $\sim r^{-1/2}\exp(\bq\br-qr)$ at large $r$. Detailed form of the density distribution $\delta n$ around a single circular obstacle have been considered in \cite{kliushnychenko_blockade_2014}. The leading asymptotics of the dissipative force and its comparison with numerical results for Eq.~(\ref{balance+}) in 2D are given in Appendix~\ref{A2}.

In the particular case of half filling ($n_0=1/2$), $\bq=0$ and the potential $G$ degenerates into usual Coulomb one, see \cite{kliushnychenko_blockade_2014} and Appendix~\ref{A2}. The form of the interaction between obstacles corresponds to anti-Newtonian dipole-dipole one as it is in the case of the linear response \cite{linear_response}. Note that single-layer potential approach enables not only correct description of the dipole-dipole interaction but also accounting for the higher order multipole moments.

The linear flow approximation allows us to describe the concentration-dependent switching of dissipative interaction between obstacles and to determine the type of this interaction. The latter, being formally expressed by Eqs.~(\ref{eq:dforcekj}) or (\ref{eq:f2}), belongs to the induced dipole-dipole (generally multipole) type of interaction in the non-equilibrium steady-state case. In contrast to usual electrostatic interaction between polarizable particles in electric field, Eqs.~(\ref{eq:dforcekj}) and (\ref{eq:f2}) describe the interaction between induced ``asymmetric'' dipoles (with nonzero total induced ``charge'', see Appendix~\ref{A2}), that is associated with anisotropic screened Coulomb-like potential with preferential direction of anisotropy $\bq$. This approximation is valid for small and widely separated obstacles and does not describe the non-linear effects that are essential for closely located obstacles as well as in the vicinity of a large-sized obstacle surfaces.

\subsection{\label{sec:nl} Nonlinear blockade effect near surface of big obstacle}

\begin{figure}
\includegraphics[width=.9\columnwidth]{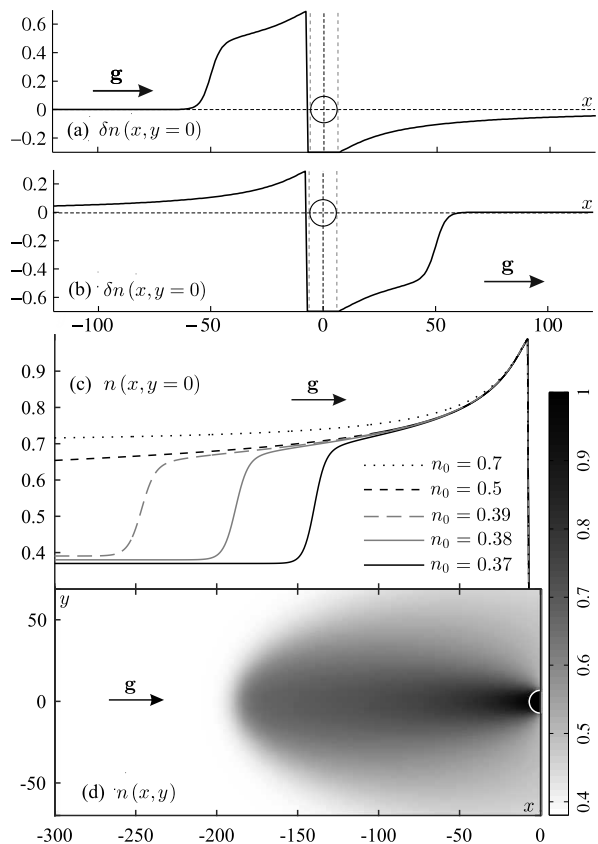}
\caption{\label{fig2}
Stationary wakes (numerical results) with kink-like profiles $\delta n(x,y=0)$ that describe cavities (a) ahead of and (b) behind the obstacle; $n_0=0.3$ for (a) and $n_0=0.7$ for (b). (c) Concentration profiles $n(x,y=0;n_0)$ at several values of the equilibrium concentration $n_0$. Note that a compact jammed region grows ahead of the obstacle as $n_0$ tends to $1/2$, for $n_0$ exceeding the half filling the wake profile becomes inverted. (d) Contour plot of concentration distribution $\delta n(x,y)=n(x,y)-n_0$ for $n_0=0.38$. Here, $|\bg|=0.5$, $\bar u=1$, $a=7$ (in units of $\ell$), and spatial coordinates are in units of $\ell$.}
\end{figure}

Here, we briefly discuss the non-linear effects caused by the gas particles blockade resorting to the numerical stationary solution of two-dimensional equation (\ref{balance+}) in the mean-field approximation.
For a relatively large obstacle, whose size is much larger than the lattice constant, the screening of the gas flow near the obstacle surface leads to a growth of the obstacle's effective size. As a result, a compact high-density (jellium-like) region is formed ahead of the obstacle, Figs.~\ref{fig2}(a), (c), and (d).
Figure~\ref{fig2}(a) shows that behavior of $\delta n$ near the obstacle surface has a pronounced step-like character at $n_0<1/2$. Note that such a step-like behavior of the density perturbation $\delta n$ is quite expected since the general type of equation (\ref{eq:lw}) admits kink-like solutions, e.g., for a one-component lattice gas, $u\equiv 0$. As gas concentration $n_0$ approaches $1/2$, the compact dense region grows (while its boundary becomes diffused), see Fig.~\ref{fig2}(c), until the uniformly decreasing distribution is formed. Further, at $n_0>1/2$, the upstream part of the profile transforms into an inverted diffusive wake with an extended dense region ahead of the obstacle, Fig.~\ref{fig2}(c), while a localized low-density region (resembling the form of a cavity) with an inverted step-like profile is formed downstream, Fig.~\ref{fig2}(b). Note that a similar compact structure occurs in a dusty plasma \cite{tsytovich_nonlinear_2013,tsytovich_self_2015}. That structure is formed by a flow of smaller dust grains ahead of a void formed by larger grains.

\begin{figure}
\includegraphics[width=.95\columnwidth]{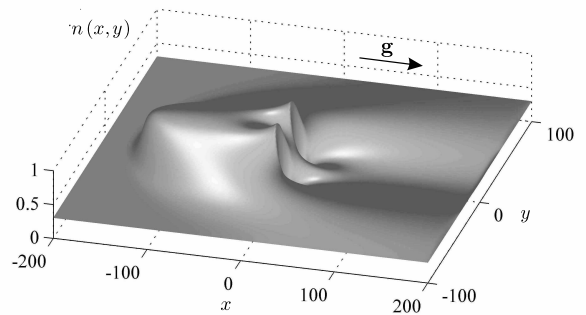}
\caption{\label{fig:ccoat}
Kink-like concentration profile $n(x,y)$ formed ahead of two closely located obstacles at $n_0=0.3$. The distance between the obstacle centers equals $4a$, other system parameters are the same as in Fig.~\ref{fig2}.}
\end{figure}

A similar nonlinear effect occurs for closely located inclusions when their individual density perturbation coats are overlapped. The overlapping leads to an additional screening of the gas flow and to the formation of a common non-linear coat around them with a step-like behavior of the density perturbation profile, Fig.~\ref{fig:ccoat}, at least in 2D case.
Note that formation of a common coat can signify the effective pairing between the inclusions (at $n_0>1/2$), i.e., formation of a stable coupled doublet \cite{mejia-monasterio_bias-_2011} (see also subsec.~\ref{sec:dfswitch}).

\subsection{\label{sec:dfswitch} Dissipative force switching for two moderately separated big obstacles}

We next consider numerically the wake-mediated force between two obstacles for two orientations of the line of their centers --- parallel and perpendicular to the gas flow. We use Eq.~(\ref{balance+}) in the mean-field approximation that takes into account the non-linear blockade effect for gas particles.
The total force exerted on a given obstacle includes the part associated with the individual friction force and the one associated with the influence of another obstacle. To separate out the inter-obstacle contribution from the total dissipative force we consider the quantity \cite{dzubiella_depletion_2003}
\begin{equation}\label{df}
  \bff_{ij}=\bff_i-\bff_i^0=\int\left[\delta n(\br,\bR_i,\bR_j)-\delta n(\br,\bR_i)\right]\nabla u_i(\br)\,\id\br,
\end{equation}
where $\bff_i$ is the total force acting on the $i$th obstacle in the presence of the $j$th one and $\bff_i^0$ is its individual friction force.

\textit{Transverse alignment} (Fig.~\ref{bconf}).
\begin{figure}
\includegraphics[width=.88\columnwidth]{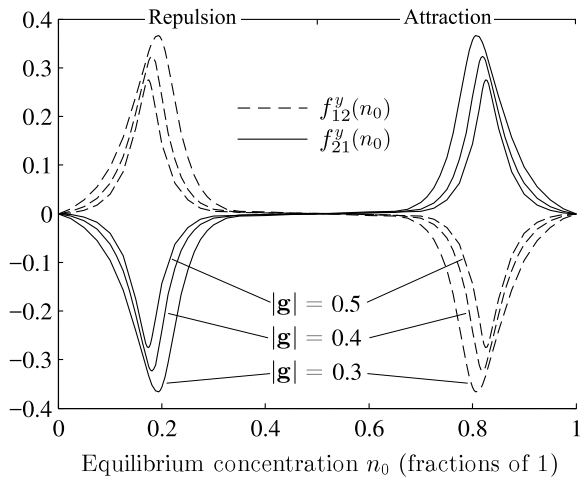}
\caption{
\label{dfswitchpconfig}
\textit{Transverse alignment}.
Dependencies of $y$-components of dissipative forces $\bff_{12}$ and $\bff_{21}$ against the equilibrium gas concentration $n_0$. Several regimes of the drive $|\bg|=0.3; 0.4; 0.5$ are plotted for comparison. Other system parameters are the same as in Fig.~\ref{bconf}, the forces are in units of $kT/\ell$ ($\ell$ is the lattice constant).
}
\end{figure}
From the symmetry of this configuration it follows that the $y$-components of the forces two obstacles exert on each other are equal and opposite, $f_{12}^y=-f_{21}^y$. At low equilibrium concentrations ($n_0<1/2$), Fig.~\ref{bconf}(a), the dissipative interaction manifests itself as an effective repulsion between the obstacles, since $f^y_{21}<0$ and $f^y_{12}>0$, see Fig.~\ref{dfswitchpconfig}. Qualitatively, this effective repulsion is simply explained by the overlap of the density coats around the obstacles that leads to formation of a dense region between them acting like a repulsive barrier, see Fig.~\ref{bconf}(a). In contrast, at $n_0>1/2$, the overlap of the individual density perturbation coats of the obstacles results in formation of an extended dense zone ahead of them that blocks the gas flow, so that the region between the obstacles becomes depleted. As Fig.~\ref{dfswitchpconfig} suggests, this collective blockade effect of gas particles leads to effective attraction between obstacles in a dense medium, $f^y_{21}>0$ and $f^y_{12}<0$.

Thus, when gas concentration $n_0$ increases, the dissipative interaction between the obstacles switches from effective repulsion to attraction. In the $n_0=1/2$ case, the effective interaction between the inclusions vanishes, $f^y_{12}=f^y_{21}=0$, regardless of the distance between them. The dissipative interaction between the inclusions naturally vanishes in the limit of empty medium $n_0\rightarrow0$, due to wake depletion. The same is true in the total jamming limit $n_0\rightarrow1$.

\begin{figure}
\includegraphics[width=.95\columnwidth]{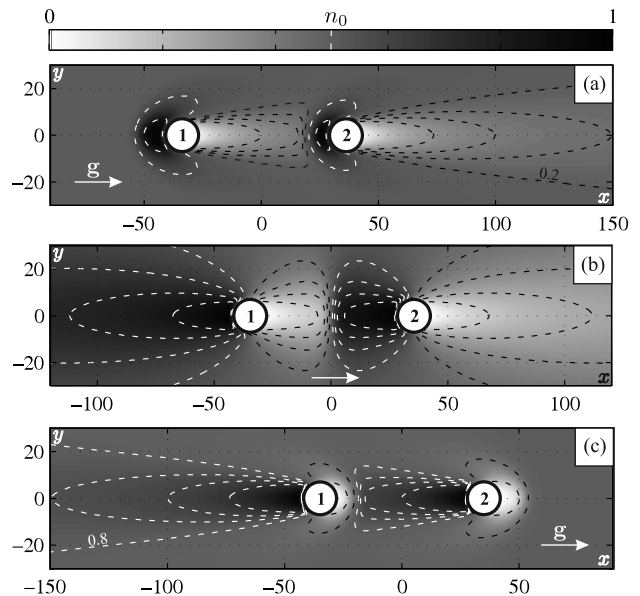}
\caption{\label{aconf} \textit{Longitudinal alignment.}
Steady-state concentration distributions (average occupation numbers) $n(x,y)$ of the gas particles near the obstacles, evaluated numerically within the mean-field approximation of Eq.~(\ref{balance+}), corresponding to the equilibrium concentrations $n_0=0.2$ (a), $0.5$ (b), and $0.8$ (c). Panels (a) to (c) illustrate three different regimes of the dissipative interaction, see Fig.~\ref{fig:afswitch}: (a) effective attraction ($|f^x_{21}|>|f^x_{12}|$), (b) anti-Newtonian interaction ($f^x_{21}=f^x_{12}$), (c) effective repulsion ($|f^x_{21}|<|f^x_{12}|$). The external field $\bg$ ($|\bg|=0.5$) is directed along the $x$-axis; the impermeable ($\bar u=1$) circular obstacles are of radius $a=7$ (in units of $\ell$), their positions are marked with the black circles; the distance between the obstacles' centers equals $10a$. The gray background corresponds to the equilibrium gas concentration $n_0$ for every contour plot, in consistence with the colorbar; spatial coordinates are in units of $\ell$.}
\end{figure}

\begin{figure}
\includegraphics[width=.9\columnwidth]{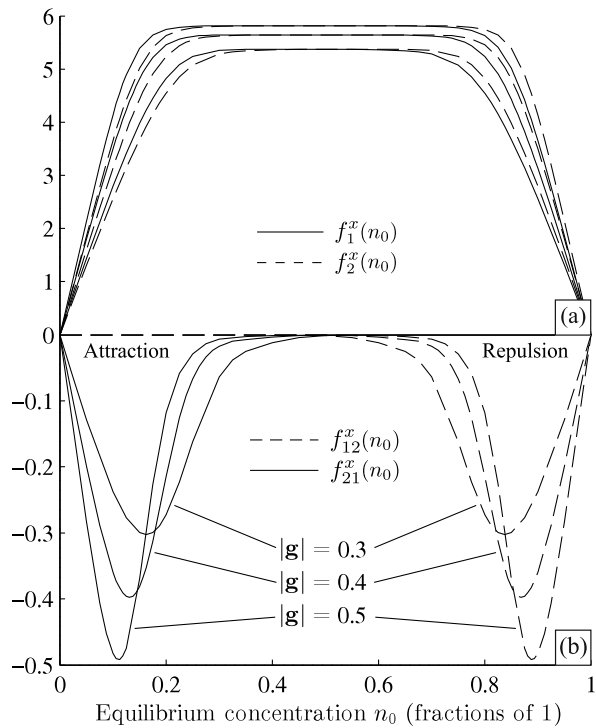}
\caption{\label{fig:afswitch}
\textit{Longitudinal alignment.}
Concentration dependence of the total forces $f^x_1(n_0)$ and $f^x_2(n_0)$ acting on each obstacle (a), and the forces $f^x_{12}(n_0)$ and $f^x_{21}(n_0)$, acting between the obstacles (b), at three magnitudes $|\bg|=0.3, 0.4, 0.5$ of the drive field. Other system parameters are the same as in Fig.~\ref{aconf}, forces are in units of $kT/\ell$ ($\ell$ is the lattice constant).
}
\end{figure}

\textit{Longitudinal alignment} (Fig.~\ref{aconf}).
At low concentrations ($n_0<1/2$), a typical situation for Brownian systems takes place: An inclusion falling on the depleted wake induced by another inclusion is effectively attracted to it since the friction force in depleted regions is weaker \cite{dzubiella_depletion_2003,khair_motion_2007}. This type of effective interaction is often referred to as the wake-mediated \cite{cividini_wake-mediated_2013,bartnick_2015}. As Fig.~\ref{fig:afswitch}(b)
suggests, the second obstacle does not practically affect the first one, $f^x_{12}\approx0$.
In contrast, at high concentrations ($n_0>1/2$), the second obstacle does not feel the influence of the first one, $f^x_{21}\approx0$, whereas the first obstacle comes under the excess pressure of the dense gas region created ahead of the second one due to the blockade effect. As a result, the effective interaction changes its sign, switching from effective attraction to repulsion, Fig.~\ref{fig:afswitch}(b). In the case of $n_0=1/2$, the effective interaction between the inclusions becomes strictly anti-Newtonian, $f^x_{12}=f^x_{21}\not=0$, see Fig.~\ref{fig:afswitch}(b). Note that for a dense gas in the blockade regime, the second obstacle ``pushes'' the first one upstream, thus reducing the total friction force $f^x_1$ exerted on the first obstacle, Fig.~\ref{fig:afswitch}(a).

The above described behavior of forces, Fig.~\ref{fig:afswitch}, can be qualitatively explained by using the results of the linear flow approximation. For example, for point-like obstacles at $n_0<1/2$, forces $f^x_{12}$ and $f^x_{21}$, see Eq.~(\ref{eq:dforcekj}), are associated with potentials $\propto\exp(-2qr_{12})/\sqrt{r_{12}}$ and $\propto1/\sqrt{r_{21}}$, respectively [see asymptotic expression (\ref{eq:A1:2dforcen}) in Appendix~\ref{A1}], so that $|f^x_{12}|\ll|f^x_{21}|$. Besides, the single-layer potential method gives correct leading asymptotics $f_{12}^x\sim |\br_{12}|^{-3/2}$ at large $|\br_{12}|$, that is in satisfactory agreement with numerical result for the general non-linear problem, Eq.~(\ref{balance+}), see Appendix~\ref{A2}.

Note that for closely located obstacles the non-linear inter-obstacle attraction can determine the dissipative pairing by the creation of common perturbation coat around them. The effect of a similar nature was obtained earlier in \cite{benichou_biased_2013} for two driven tracers. Indeed, at high gas concentration ($n_0>1/2$) the depleted cavities formed around each obstacle, see Figs.~\ref{bconf}(c) or \ref{aconf}(c), can entail specific behavior of dissipative forces depending on the distance between obstacles. In particular, the effective interaction between two obstacles in close proximity undergoes an abrupt change in the asymptotic behavior, see Appendix~\ref{A2} and figures therein, that can be indicative of the dissipative pairing effect.
\begin{figure}
\includegraphics[width=.85\columnwidth]{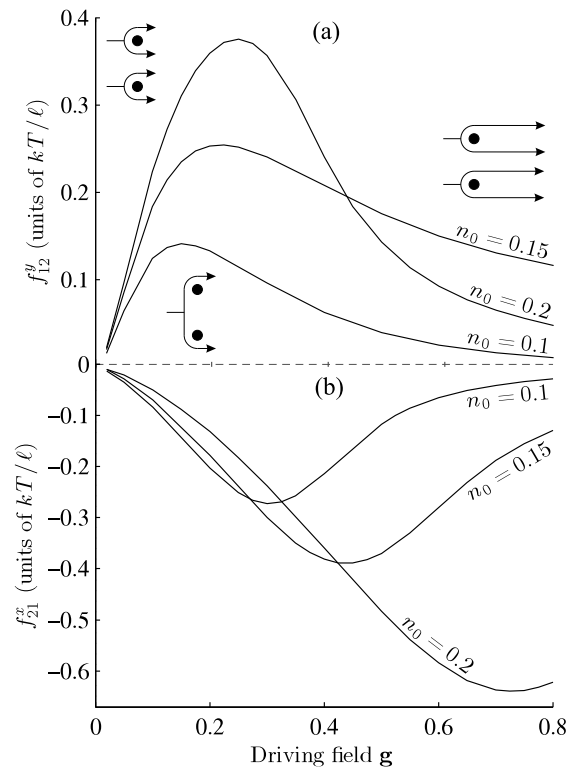}
\caption{\label{fig:drive}
Dissipative forces (a) $f_{21}^x$ (for longitudinal alignment) and (b) $f_{21}^y$ (for transverse one) versus the external drive $\bg$ for the gas concentrations $n_0=0.1; 0.2; 0.3$. Obstacle sizes and separation correspond to that on Figs.~\ref{bconf} and \ref{aconf}. Schematic illustrations of the shape transformations of density perturbation coats are shown for transverse alignment.
}
\end{figure}

Dependence of the strength of dissipative interaction between obstacles on the external driving field $\bg$ appears to be non-monotonic, see Fig.~\ref{fig:drive}. The characteristic peak of interaction corresponds to the drive magnitude when the most efficient common density coat is formed around the obstacle pair. This behavior can be explained by the changes in the shape of density perturbations, e.g., for the case $0<n_0<1/2$, see Fig.~\ref{fig:drive}(a). At low gas concentration the effective repulsion between obstacles vanishes in the limit of weak driving, since slow flow of a sparse gas does not induce significant gas perturbations and, thus, wake-mediated interactions. The characteristic peak of interaction corresponds to the driving magnitude when the common density coat is formed (see schematic illustrations on Fig.~\ref{fig:drive}): in this regime, profile of the density perturbation provides the most efficient dissipative wake-mediated influence between obstacles. Strong driving field causes the perturbation coat around each obstacle to decrease in lateral dimension and increase in longitudinal, so that overall density coat stretches along the flow direction. As a result, overlap of the individual obstacles' coats reduces, and their mutual influence decays. In other words, strong enough drift flow reduces the common density coat. This qualitative reasoning is also true in the case of effective attraction under longitudinal alignment. Note that the peak position shifts and increases towards the region of strong driving as gas concentration decreases in case of longitudinal alignment of obstacles, Fig.~\ref{fig:drive}(b), while in case of transverse one the situation is just the opposite, Fig.~\ref{fig:drive}(a). Hence, the most favorable condition for the pronounced common coat organization is determined by both the equilibrium gas concentration $n_0$ and the strength of external driving field $\bg$.

The magnitude of the evaluated forces can be easily estimated, e.g., for the case of atoms adsorbed on solid surface. Choosing the lattice spacing parameter to be $\ell=3$ {\AA}, at room temperature one obtains the range of dissipative forces to be 5--10 pN [see Fig.~\ref{fig:afswitch}(b)], while the friction force is approximately one order of magnitude stronger [see Fig.~\ref{fig:afswitch}(a)]. Notice that the same ratio between magnitudes of friction and dissipative forces is observed for probe-colloids moving through a colloidal suspension in 3D case \cite{sriram_out--equilibrium_2012,sriram_two_spheres_2015}. In addition, as is seen from Fig.~\ref{fig:afswitch}(a), at concentrations close to $n_0=1/2$ the forces exerted on each obstacle by the gas are almost equal, i.e., the dissipative interaction between obstacles takes anti-Newtonian character, $f^x_{12}=f^x_{21}$, see Fig.~\ref{fig:afswitch}(b). It should be mentioned that an analogous behavior occurs for two probes moving along their line of centers through a colloidal suspension: both inclusions may experience the same drag force, as was observed in a recent experiment \cite{sriram_two_spheres_2015}, at the effective volume fraction of $0.41$. However, in this case, the effect is due to the hydrodynamic interactions between bath particles.

\section{\label{sec:conclusion} Conclusion}

Let us briefly summarize the main obtained results for dissipative (wake-mediated) interaction between obstacles embedded into gas flow with taking into account short-range repulsive interaction between gas particles.

--- We have shown that increasing of the gas concentration enhances the role of inter-particle interaction and can lead to the sign change of the effective interaction between obstacles, i.e., switching from effective attraction to repulsion or vice-versa. The effect of concentration-dependent force switching is associated with obstacle wake transformation --- its inversion.

--- In the case of small and widely separated obstacles, the wake-mediated dissipative interaction between them has been shown to belong to the type of induced dipole-dipole (generally, multipole) interaction associated with anisotropic screened Coulomb potential. To this end, we have developed the representation for the gas density perturbation in the form of single-layer potential. Formally, this is a generalization of the single-layer potential approach for the electrostatic interaction between polarizable particles induced by stationary external field. Our approach is applicable to non-equilibrium steady-state case where interaction between the obstacles is induced by gas flow. Obtained analytical expressions qualitatively explain the asymmetry of the obstacle's wake, the long-range behavior of dissipative interaction, its non-Newtonian character, and switching of both the wake direction and the dissipative forces.

--- Dissipative interaction between obstacles is most pronounced when a common perturbation coat around them (collective wake) is formed. The force depends non-monotonically on equilibrium gas concentration, magnitude of external sweeping field (gas flow), and alignment of the obstacles. In particular, at low gas concentrations two obstacles are effectively attracted in the case of longitudinal alignment and repel each other in the case of transverse one. At high gas concentrations the situation is just the opposite.

--- The non-linear blockade effect of gas particles is significant near the surface of relatively big obstacles and/or for closely located ones. In this case, repulsive interaction between gas particles has been shown to lead to screening of the gas flow near the obstacles and to formation of a common coat of gas density perturbation around them, with pronounced step-like behavior of the density profile. Formation of common coat can determine the non-linear mechanism of dissipative pairing between the obstacles (see, e.g., \cite{tahir-kheli_correlated_1983,mejia-monasterio_bias-_2011}).

It should be noted that we initially used rough approximations, so that a number of important questions were left behind the scope of our paper. In particular, using the mean-field approximation, we lose information on the short-range correlations in a gas, such as ``back correlations'', see, e.g., \cite{tahir-kheli_correlated_1983,mejia-monasterio_bias-_2011}, which have to occur near the obstacle surfaces (another gas component). In addition, neglecting fluctuations in a gas, i.e., the term $\delta J_i^\alpha$ in Eq.~(\ref{balance+}), we do not take into account the fluctuation-induced (Casimir-like) forces, see, e.g., \cite{bitbol_forces_2011,demery_thermal_2011,bartolo_fluctuations_2002,dean_out--equilibrium_2010,krech_fluctuation_1999,Buzzaccaro_Critical_2010,Piazza_Critical_2011} which can be significant for pairing effect at small inter-obstacle distance.

Obtained results may be of interest when considering the dissipative structure formation (see, e.g., \cite{bartnick_2015}), collective friction force or collective energy losses in an ensemble of inclusions, and can find applications in systems with driven hopping transport (e.g., surface kinetics of adsorbed atoms \cite{chumak_diffusion_1980,chumak_1999,benichou_stokes_2000,benichou_force_2001}, fast ionic conductors, etc.) or serve as a rough model for colloidal suspensions or dusty/complex plasma \cite{tsytovich_nonlinear_2013,tsytovich_self_2015}.

\acknowledgments{
We are grateful to A.~A.~Chumak, B.~I.~Lev, V.~V.~Gozhenko, and V.~V.~Bondarenko for helpful discussions and comments on the manuscript.}

\appendix
\section{\label{A1} Qualitative picture of dissipative interaction: 
a rough analytical approach
}

In this appendix, we roughly estimate the wake-mediated interaction between widely separated small obstacles imbedded into gas flow. Let us consider a non-equilibrium steady-state problem in the long-wavelength approximation, Eq.~(\ref{eq:steady}):
\begin{equation}\label{eq:A1:01}
  \nabla^2 n - U\nabla^2 n + n\nabla^2 U - (\bg,\nabla)n(1 - n - U)=0,
\end{equation}
where inclusions are given by a distribution $U$ of the heavy gas component. For simplicity, we consider a smooth distribution $U(\br)=\sum_k u(\br-\bR_k)$, where distribution $u(\br-\bR_k)$ describes $k$th inclusion and has a compact carrier located near the inclusion center $\bR_k$. For distant inclusions, we assume that $\int u(\br-\bR_k)u(\br-\bR_j)\,\id \br \approx 0$.

For widely separated small obstacles (whose sizes are comparable with the lattice constant), interaction between the particles is less significant and the drift term in Eq.~(\ref{eq:A1:01}) can be written in the linear approximation, see \cite{kliushnychenko_blockade_2014}. Assuming that distribution $n=n_0+\delta n$ weakly deviates from the equilibrium one $n_0$, we linearize the drift flow term in Eq.~(\ref{eq:A1:01}), taking $n^2\approx n_0^2 + 2n_0 \delta n$, and rewrite the equation in the following form:
\begin{align}\label{eq:A1:01.5}
  &\nabla^2\delta n - 2(\bq,\nabla)\delta n\nonumber\\
 = {}&U\nabla^2\delta n - (n_0+\delta n)\nabla^2U - (\bg,\nabla)(n_0+\delta n)U,
\end{align}
where $\bq=(1/2-n_0)\bg$. Based on Eq.~(\ref{eq:A1:01.5}), we estimate the asymptotic behavior of the dissipative interaction between the obstacles depending on the distance between them, their mutual alignment, and equilibrium gas concentration $n_0$. We shall use a qualitative approach which allows us to obtain simple analytical expressions for density perturbation and dissipative forces.

It is convenient to consider an integral representation of Eq.~(\ref{eq:A1:01}) using the Green's function $G(\br-\br')$ of the equation
\begin{equation}\label{eq:A1:02}
  \nabla^2_\br G(\br - \br') - 2(\bq,\nabla_\br) G(\br-\br') = -\delta(\br - \br').
\end{equation}
The form of this Green's function is similar to the anisotropic screened Coulomb potential
\begin{equation}\label{eq:A1:3G}
    G(\br-\br')=\frac{e^{-q|\br-\br'|+\bq(\br-\br')}}{4\pi |\br-\br'|}
\end{equation}
in 3D case and
\begin{equation}\label{eq:A1:2G}
    G(\br-\br')=\frac{1}{2\pi}e^{\bq(\br-\br')}K_0(q|\br-\br'|)
\end{equation}
in 2D case. By using (\ref{eq:A1:02}), we can rewrite the equation for the gas density perturbation $\delta n$ in the form
\begin{align}\label{eq:A1:07}
    \delta n (\br) = {}&[n_0 + \delta n(\br)]U(\br)\nonumber\\
    &+ \int U(\br')\left(\bOmega(\br'),\nabla_\br\right) G(\br - \br')\,\id\br',
\end{align}
where
\begin{equation}\label{eq:A1:08}
  \bOmega(\br) = 2\bg (1-n_0)[n_0+\delta n(\br)] - 2\nabla_{\br}\delta n(\br).
\end{equation}
Equation (\ref{eq:A1:07}) can be simplified by applying an approach similar to the self-consistent molecular field approach \cite{Lev_Interaction_1999}. Since distribution $U$ of the heavy component is localized near the inclusion centers $\bR_j$ and has compact carriers $u_j(\br)=u(\br-\bR_j)\leq1$, we may consider $u_j(\br)$ as a probability density distribution and the integral in (\ref{eq:A1:07}) as an average $\overline {(\bOmega , \nabla G)}_j$ associated with this distribution. Here, $\overline{(\ldots)}_j=\int(\ldots)u(\br-\bR_j)\,\id\br$. Then, using the mean-field approximation, $\overline {(\bOmega , \nabla G)}_j\approx\left(\overline{\bOmega}_j,\overline{\nabla G}_j\right)\approx \left(\overline{\bOmega}_j,\nabla G(\br-\bR_j)\right)$, Eq.~(\ref{eq:A1:07}) can be rewritten as
\begin{equation}\label{eq:A1:6*}
  \delta n(\br)\approx [n_0+\delta n(\br)]U(\br) + \sum_j\left(\overline{\bOmega(\bR_j)},\nabla_\br\right)G(\br-\bR_j),
\end{equation}
where
\begin{equation}\label{eq:A1:7**}
  \overline{\bOmega(\bR_j)} = 2\bg(1-n_0)\left[n_0 + \overline{\delta n_j}\right] - 2\overline{\nabla \delta n_j}
\end{equation}
plays the role of a molecular field or an average flux in the system, these quantities being defined by external field $\bg$ and the density perturbation field due to other inclusions. Equations for the constants $\overline{\delta n}_j$ and $\overline{\nabla \delta n}_j$ can be obtained in a self-consistent manner by using Eq.~(\ref{eq:A1:6*}), see Ref.~\cite{a}.

Representation (\ref{eq:A1:08}) enables us to estimate qualitatively the asymptotic behavior of gas density perturbation far from an isolated inclusion and the asymptotic behavior of the dissipative force between widely separated inclusions. Using (\ref{eq:A1:3G}), gas density perturbation (\ref{eq:A1:6*}) far from an isolated inclusion can be written as
\begin{equation}\label{eq:A1:deltaen}
    \delta n(\br) \sim \left(\overline{\bOmega},\nabla_{\br} \right)G(\br),
\end{equation}
that is
\begin{equation}
  \delta n(\br) \sim \left(\overline{\bOmega},\nabla_{\br}\right)
  \frac{e^{-q|\br|+\bq\br}}{|\br|}
\end{equation}
in 3D case and
\begin{equation}
  \delta n(\br) \sim  \left(\overline{\bOmega},\nabla_{\br}\right)
  \frac{e^{-q|\br|+\bq\br}}{|\br|^{-1/2}}
\end{equation}
in 2D case. In the last expression, the asymptotic behavior of the Bessel function $K_0(q|\br|) \sim |\br|^{-1/2}e^{-q|\br|}$ for large $r$ was used. At low concentrations of gas ($n_0<1/2$), the dense region ahead of the inclusion is described by an exponential asymptotics, while the asymptotics of the depleted region behind the inclusion is power-law. When gas concentration increases and $n_0$ becomes greater than $1/2$, vector $\bq=(1/2-n_0)\bg$ changes its direction. It means that switching of the wake direction occurs, together with corresponding switching between the exponential and power-law asymptotics. At $n_0=1/2$ we have $\bq=0$ and the asymptotics of perturbation $\delta n$ corresponds to a dipole-like polarization of gas density perturbation around the inclusion.

The force acting on the $k$th inclusion is $\bff_k = \int \delta n(\br)\nabla _{\br}u(\br-\bR_k)\,\id\br \,\sim -\overline {\nabla \delta n_k}$. For small (point-like) inclusions, the force exerted by the $j$th inclusion on the $k$th one can be roughly estimated as
\begin{equation}\label{eq:A1:dforcekj}
\bff_{kj} \sim  -\nabla_{\bR_k}\left(\overline\bOmega_j,\nabla_{\bR_k}\right)G(\br),
\end{equation}
that is
\begin{equation}
  \bff_{kj} \sim  -\nabla_{\bR_k}\left(\overline\bOmega_j,\nabla_{\bR_k}\right) \frac{e^{-q|\bR_k-\bR_j|+\bq(\bR_k-\bR_j)}}{|\bR_k-\bR_j|}
\end{equation}
in 3D case and
\begin{equation}\label{eq:A1:2dforcen}
  \bff_{kj} \sim  -\nabla_{\bR_k}\left(\overline\bOmega_j, \nabla_{\bR_k}\right) \frac{e^{-q|\bR_k-\bR_j|+\bq(\bR_k-\bR_j)}}{|\bR_k-\bR_j|^{-1/2}},
\end{equation}
in 2D. As is seen from Eqs.~(\ref{eq:A1:deltaen}) and (\ref{eq:A1:dforcekj}), the local density perturbation around an obstacle is formed by an effective flow $\overline \bOmega$ (molecular field) that is determined by the external flow and the flows induced by gas density perturbations of all the inclusions.

\section{\label{A2} Single-layer potential approach}

A more rigorous result for wakes and for the dissipative force can be obtained in the framework of the single-layer potential method for inclusions with sharp boundaries. Again, we start from Eq.~(\ref{eq:steady}). It is convenient to use a new function $\psi(\br) = n(\br)/[1-u(\br)]$ governed by the equation (see \cite{kliushnychenko_blockade_2014})
\begin{equation}\label{eq:A2:psieq}
  \nabla\left\{\varepsilon \left[\nabla\psi-\psi\left(1-\psi\right)\bg\right]\right\}=0,
\end{equation}
where $\varepsilon=\varepsilon(\br)=[1-u(\br)]^{2}$, and it is assumed that $u(\br)\neq 1$. Let us represent the solution $\psi(\br)\approx\psi_0+\delta\psi(\br)$ as a small deviation $\delta\psi(\br)$ from the equilibrium distribution $\psi_0\equiv n_0$, and linearize Eq.~(\ref{eq:A2:psieq}):
\begin{equation}\label{eq:A2:psieqLinear0}
  \nabla \left[\varepsilon\left(\nabla\delta\psi-2\bq
  \delta\psi-\bQ\right)\right]=0,
\end{equation}
where $\bQ=n_0(1-n_0)\bg$ and $\bq=(1/2-n_0)\bg$. This linear equation takes into account the interaction between gas particles in the first order of the perturbation theory. In this sense, (\ref{eq:A2:psieqLinear0}) is the simplest possible generalization of the drift-diffusion equation that was exploited in \cite{dzubiella_depletion_2003} for a gas of non-interacting particles at low concentrations.

The inclusions are represented by the distributions of heavy gas-component $u_j(\br)=u(\br-\bR_j)$ centered at points $\bR_j$ with homogeneous concentration $u_j(\br)=\bar u_j=\mathrm{const}$ inside inclusions and $u_j(\br)\equiv0$ outside them. Note that in the case of inclusions with sharp boundaries, Eq.~(\ref{eq:A2:psieq}) allows for a solution in the class of continuous functions, whereas function $n(\br)$, obeying Eq.~(\ref{eq:A1:01}), as well as its normal derivative have a jump at the inclusion's boundary. The density perturbation inside ($\delta\psi^-$) and outside ($\delta\psi^+\equiv\delta n$) the inclusions obey the equation
\begin{equation}\label{eq:A2:psieqLinear}
  \nabla \left(\nabla\delta\psi^\pm-2\bq
  \delta\psi^\pm-\bQ\right)=0.
\end{equation}
Equation (\ref{eq:A2:psieqLinear}) is supplemented by the matching conditions for $\delta\psi^\pm$ on the surface $S_i$ of $i$th inclusion:
\begin{align} \label{eq:A2:bcPSI}
  &\delta\psi^+(\br)=\delta\psi^-(\br),\nonumber\\
  &\varepsilon^+\left[\nabla_\bn^+\delta\psi^+(\br)-2q_n\delta\psi^+(\br)-Q_n\right]\nonumber\\
 =&\varepsilon^-_i\left[\nabla_\bn^-\delta\psi^-(\br)-2q_n\delta\psi^-(\br)-Q_n\right],
  \end{align}
where $Q_n=(\bQ,\bn_{\br})$, $q_n=(\bq,\bn_{\br})$, $\bn_\br$ is the outward normal at the point $\br\in S_i$, $\varepsilon^+=1$ outside the inclusions and $\varepsilon^-_i=(1-\bar u_i)^2$ inside the $i$th inclusion, and notation $\nabla_\bn^\pm(\ldots)\equiv\lim_{|\widetilde{\br}-\bR_i|\rightarrow |\br-\bR_i|\pm0}\left(\frac{\partial(\ldots)}{\partial\bn}\right)(\widetilde{\br})$ is used.

The solution of (\ref{eq:A2:psieqLinear})--(\ref{eq:A2:bcPSI}) can be represented in the form of a single-layer potential, similarly to that used in \cite{kliushnychenko_blockade_2014} for a single obstacle,
\begin{equation}\label{eq:A2:slp}
  \delta\psi(\br) = \sum_i \int_{S_i} G(\br-\br')\mu_i(\br')\,\id \br',
\end{equation}
where $G(\br-\br')$ is the Green's function, (\ref{eq:A1:3G}) in 3D, or (\ref{eq:A1:2G}) in 2D. The quantity $\mu_i(\br')$ plays the role of a ``charge'' density induced by the external field $\bg$ on the obstacle surface $S_i$ \cite{mayergoyz_2005}. It satisfies the following integral equation determined by the matching conditions (\ref{eq:A2:bcPSI}):
\begin{gather}
    2\lambda_i\left[\nabla_\bn^+ -2q_n(\br_i)\right]\sum_j \int_{S_j} G(\br_i - \br_j)\mu_j (\br_j)\,\id \br_j \nonumber\\
    +(\lambda_i-1)\mu_i(\br_i)=2\lambda_i Q_n(\br_i),\label{eq:A2:intGeen}
\end{gather}
where $\br_i \in S_i$ and $\lambda_i=\lambda(\bar u_i)=(\varepsilon^+-\varepsilon^-_i)/
(\varepsilon^++\varepsilon^-_i)$. Equation (\ref{eq:A2:intGeen}) was derived with the use of the jump theorem for the normal derivative of the potential of a single layer on an obstacle surface \cite{vladimirov_1985}, $\nabla_\bn^\pm \delta\psi^\pm(\br)=\mp\mu(\br)/2 + \nabla_\bn\delta\psi(\br)$.
Representation (\ref{eq:A2:slp}) and Eq.~(\ref{eq:A2:intGeen}) describe the general solution for obstacles with arbitrary geometry of their surfaces (Lyapunov surface, see \cite{vladimirov_1985}).

Considering that $\delta \psi^+(\br)\equiv\delta n(\br)$ and using Eq.~(\ref{eq:force}), we can write final expression for the density perturbation around obstacles and the force acting on an obstacle, both being induced by sweeping field $\bg$ (or by the gas flow):
\begin{equation}
  \delta n(\br) = \sum_j\int_{S_j}G(\br-\br_j)\mu_j(\br_j)\,\id \br_j,
\end{equation}
\begin{equation}
  \bff_k = -\sum_j\int_{S_k}\int_{S_j}\bn(\br_k)G(\br_k-\br_j)\mu_j(\br_j)\,\id\br_k \id\br_j.
\end{equation}
This representation of the solution has direct analogy with induced interaction between dielectric particles in a stationary electric field $\bQ$: External electric field induces charge $\mu$ on the particle surface, leading to its polarization, e.g., inducing the dipole moment for a spherical particle, see \cite{landau_electrodynamics_1984}. This, in turn, leads to multipole (e.g., dipole-dipole) interaction between the particles. However, in our case, contrary to the electrostatic problem the density $\mu_k$ is induced by an external field (flow) on the inclusion surfaces, and multipole interaction between them is determined not by the Coulomb potential $|\br|^{-1}$ but anisotropic screened Coulomb-like potential $|\br|^{-1}\exp(\bq\br-q|\br|)$ (in 3D case), see (\ref{eq:A1:3G}) and (\ref{eq:A1:2G}). Such form of the potential leads, in particular, to nonconservation of induced surface density, $\int\mu\, \id S \neq0$, and to asymmetric distribution of ``induced potential'' $\delta n$ near the inclusion. The latter describes inclusion wake, e.g., wake with a localized region of dense gas ahead of the inclusion and an extended depleted tail [see, e.g., Fig.~\ref{bconf}(a)].

In the particular case of the half filling ($n_0=1/2$), the second term in equation (\ref{eq:A2:psieqLinear0}) vanishes ($\bq\equiv0$) and the problem is reduced to an electrostatic-like problem $\nabla\left[\varepsilon(\nabla\delta\psi-\bQ)\right]=0$ for dielectric particles in a uniform electric field $\bQ = \bg/4$. In this case, density distribution $\delta\psi(\br)$ is similar to the distribution of the electrostatic potential characterizing the scattered field. It means that the induced interaction between obstacles via their common environment (density perturbation) behaves like electrostatic dipole-dipole (generally, multipole) interaction. For a single obstacle with radius $a$, density perturbation $\delta n=\delta\psi^+$ around the obstacle at $n_0=1/2$ can be obtained in an explicit form: $\delta n = \lambda a^2(\bQ,\nabla_\br)\ln a|\br|^{-1}$ for 2D case and $\delta n = \lambda a^2(\bQ,\nabla_\br)|\br|^{-1}$ for 3D. These results explain both the power-law asymptotic behavior of gas perturbation and the symmetry of the ``upstream/downstream'' tail, see Fig.~\ref{fig2}(b) and \cite{kliushnychenko_blockade_2014}. This case ($n_0=1/2$) corresponds to the linear response of $\delta n$ to the external field $\bg$, cf. \cite{linear_response}. Note that symmetry of wake (or profile of perturbation) generated in a medium by a moving probe particle is a common result for systems described in the linear response approximation (see, e.g., \cite{pines_theory_1966}).

For widely separated inclusions, when the distance between their centers $|\br_{kj}|=|\bR_k-\bR_j|$ is much larger than their characteristic sizes $a_j$, $|\br_{kj}|=|\bR_k-\bR_j|\gg a_k (a_j)$, the multipole expansion of the potential $G$ can be used:
\begin{equation}\label{eq:A2:expansion}
  G(\br_k-\br_j) \approx G(\br_{kj}) + \left(\bx_{kj}\cdot\nabla_{\br_{kj}}\right)G(\br_{kj})+\cdots,
\end{equation}
where $\bx_{kj} = \bx_k-\bx_i$ and $\bx_k = \br_k - \bR_k$.
Next we consider the particular 3D case for spherical obstacles with radii $a_k$. For obstacles located far from each other, $|\br_{ki}|\gg|\bx_{ki}|$, one can use the multipole expansion (\ref{eq:A2:expansion}) for the kernel of integral operator in (\ref{eq:A2:intGeen}). In the dipole approximation, the integral equation for the induced surface density $\mu_k (\bx_k)$ on the surface of the $k$th inclusion takes the form
\begin{widetext}
\begin{equation}\label{eq:A2:B3}
  \hat{\Lambda}_{\bx_k}\,\mu_k(\bx_k) +
  \bx_k \left(\nabla_{\bx_k} ^+ -2\bq \right)\sum_{i\neq k}\frac{e^{-q|\br_{ki}|+\bq \br_{ki}}}{4\pi|\br_{ki}|} \int_{S_i} (1+\bx_{ki} \bu_{ki})\, \mu_i (\bx_i) \, \id\bx_i \,=\, \bx_k \bQ,
\end{equation}
where
\begin{equation}
\bu_{ki}(\br_{ki})= \bq -q \frac {\br_{ki}}{|\br_{ki}|}-\frac{\br_{ki}}{|\br_{ki}|^2},
\end{equation}
and $\hat{\Lambda}_{\bx_k}$ denotes the integral operator for a single obstacle
\begin{equation}\label{eq:A2:B2}
\hat{\Lambda}_{\bx_k}\, \mu_k(\bx_k)=a_k \frac{\lambda_k -1}{2 \lambda_k}\mu_k (\bx_k)+\bx_k \left(\nabla_{\bx_k} ^+ -2\bq \right)\int_{S_k}\frac{e^{-q|\bx_{kk}|+\bq \bx_{kk}}}{4\pi|\bx_{kk}|} \, \mu_k (\bx'_k) \, \id\bx'_k.
\end{equation}
Equation (\ref{eq:A2:B3}) for $\mu_k$ has small parameter $\exp{(-q|\br_{ki}|+\bq \br_{ki})} / 4\pi|\br_{ki}|\, \ll\,1$, that allows us to consider the influence of other obstacles on a given one as a small perturbation $\mu_k ^1$ of the solution $\mu_k ^0=\mu ^0$ for a single obstacle, $\mu_k\approx\mu_k ^0+\mu_k ^1$. In this approximation, equations for $\mu_k ^0=\mu ^0$ and $\mu_k ^1$ take the form
\begin{equation}\label{eq:A2:B4}
\hat{\Lambda}_{\bx_k}\,\mu_k ^0 (\bx_k) =\, \bx_k \bQ,
\end{equation}
\begin{equation}\label{eq:A2:B5}
\hat{\Lambda}\,\mu_k ^1 (\bx_k) \,=\,
-\bx_k \left(\nabla_{\bx_k} -2\bq \right)\sum_{i\neq k}\frac{e^{-q|\br_{ki}|+\bq \br_{ki}}}{4\pi|\br_{ki}|}
\int_{S_i} (1+\bx_{ki} \bu_{ki})\, \mu_i ^0 (\bx_i) \, \id\bx_i.
\end{equation}
The formal solution of the last equation can be written as
\begin{equation}\label{eq:A2:mu1}
\mu_k ^1 (\bx_k) \,=\,
\sum_{i\neq k}\frac{e^{-q|\br_{ki}|+\bq \br_{ki}}}{4\pi|\br_{ki}|}
\hat{\Lambda}^{-1}_{\bx_k}\int_{S_i} W(\bx_k,\bx_i,\br_{ki})\, \mu_i ^0 (\bx_i) \, \id\bx_i,
\end{equation}
where
\begin{equation}
  W(\bx_k,\bx_i,\br_{ki})=2(\bq \bx_k)(1+\bx_{ki} \bu_{ki})-\bx_k \bu_{ki}.
\end{equation}
\end{widetext}

The dissipative force acting on an obstacle is determined by the gas density perturbation $\delta n(\br)$ on its surface, Eq.~(\ref{eq:force}). In this case we can set $\delta n\approx\delta \psi(\br)$, where $\br\approx \bR_k + \bx_k $ and $|\bx_k|=a_k$ is the radius of the $k$th obstacle. In the dipole approximation, Eqs.~(\ref{eq:A2:expansion}), (\ref{eq:A2:B4}), (\ref{eq:A2:B5}), density perturbation near the $k$th obstacle can be written in the form
\begin{align}\label{Neighbor_Infl}
  \delta n_k &\approx \delta n_k^0 + \int_{S_k}G(\bx_{kk})\mu_k^1(\bx_k')\,\id\bx_k' \nonumber\\
  & + \sum_{i\neq k}\int_{S_i}\left[G(\br_{ki}) + \left(\bx_{ki}\cdot\nabla_{\br_{ki}}\right)G(\br_{ki})\right]\mu_i^0(\bx_k')\,\id\bx_k'.
\end{align}
The right-hand side of expression (\ref{Neighbor_Infl}) containing the sum over all the obstacles $i\neq k$ describes their direct influence on the given $k$th obstacle. The first term in Eq.~(\ref{Neighbor_Infl})
\begin{equation}
  \delta n_k^0=\int_{S_k}G(\bx_{kk})\mu_k^0(\bx_k)\,\id\bx_k'
\end{equation}
gives the contribution to the gas perturbation around the $k$th obstacle caused by the $k$th obstacle itself.

Using Eqs.~(\ref{eq:A2:mu1}) and (\ref{Neighbor_Infl}), contribution to the density perturbation $\delta n_k$ near the $k$th inclusion caused by other inclusions in 3D case can be written as
\begin{equation}\label{eq:A2:deltaenka}
  \delta n_k - \delta n_k ^0 \approx \sum_{i\not=k}\delta n_{ki},
\end{equation}
where
\begin{equation}\label{eq:A2:Asym_Screened_Coulomb}
  \delta n_{ki} \sim \frac{e^{-q|\br_{ki}|+\bq\br_{ki}}}{4\pi|\br_{ki}|}
  I\left(\br_{ki},\bq,\bx_k\right)
\end{equation}
is contribution of the $i$th obstacle to the density perturbation near the $k$th obstacle surface,
\begin{align}\label{eq:A2:Iki}
  &I\left(\br_{ki},\bq,\bx_k\right)=
  \int_{S_i} \biggl[\left(1+\bx_{ki}\bu_{ki}\right)
  \mu^0_i(\bx'_i) \nonumber\\&
  +\int_{S_k}\frac{e^{-q|\bx_{kk}|+\bq\bx_{kk}}}{4\pi|\bx_{kk}|}
  \hat{\Lambda}^{-1}_{\bx_k '} W(\bx'_k,\bx'_i,\br_{ki})\, \mu_i ^0 (\bx'_i)\id\bx'_k\biggr] \, \id\bx'_i.
\end{align}

As it follows from Eq.~(\ref{eq:A2:Iki}), $I\left(\br_{ki},\bq,\bx_k\right)$ has a power-law dependence on $\br_{ki}$ and in the case of $a_i\ll r_{ik}$ depends only on the mutual alignment of the obstacles with respect to the external field $\bg$, i.e., on $\theta_{ki}$, the angle between $\br_{ki}$ and $\bg$.

Using expression (\ref{eq:A2:Asym_Screened_Coulomb}) for the density perturbation $\delta n_k$, we can represent the force exerted on the $k$th inclusion by the $i$th one in the form that is similar to (\ref{eq:A1:dforcekj}):
\begin{gather}
  \bff_{ki}\approx-\int_{S_k}\bn(\bx_k)\delta n_{ki}(\bx_k)\,\id\bx_k\nonumber\\
  =-\frac{e^{-q(1-\beta\cos \theta_{ki})|\br_{ki}|}}{4\pi|\br_{ki}|}
  \int_{S_k}\bn(\bx_k) I\left(\br_{ki},\bq,\bx_k\right)
  \,\id\bx_k.\label{eq:A2:f2}
\end{gather}
Expressions (\ref{eq:A2:Asym_Screened_Coulomb})--(\ref{eq:A2:f2}) are obtained in the dipole approximation and give a rough asymptotic behavior of the induced non-equilibrium correlations and dissipative forces between two obstacles located far from each other, depending on the distance between them $|\br_{ik}|$ and their mutual alignment $\theta_{ik}$ with respect to the external field $\bg$. In view of (\ref{eq:A2:Asym_Screened_Coulomb})--(\ref{eq:A2:f2}), the influence of the $i$th obstacle on the $k$th one is not equivalent to that of the $k$th obstacle on the $i$th one ($\theta_{ki}=\pi - \theta_{ik}$), i.e., these correlations are not reciprocal, $\delta n_{ki}\neq\delta n_{ik}$, and the forces are non-Newtonian, $\bff_{ki}\neq -\bff_{ik}$.

As is seen from (\ref{eq:A2:f2}), dissipative forces acting between inclusions are expressed, in the dipole approximation, in terms of induced density $\mu^0$ of isolated inclusion. Distribution $n^0(\br)=n_0+\delta n^0(\br)$ for a single obstacle in 3D case takes the form
\begin{equation}
  n^0(\br)=n_0+\int_S\frac{e^{\bq(\br-\br')-q|\br-\br'|}}{4\pi |\br-\br'|}\mu^0(\br')\,\id \br'.
\end{equation}

Far from the obstacle, when $|\br|\gg a$ ($a$ is its characteristic size), we can easily extract the leading asymptotics for the gas density perturbation $\delta n$ induced by the external field $\bq=(1/2-n_0)\bg$:
\begin{equation}\label{eq:A2:n13D}
    \delta n(\br)\approx \frac{e^{\bq\br-q|\br|}}{|\br|}\tilde I(\br,\bq),
\end{equation}
[compare with Eq.~(\ref{eq:A1:07})]. $\tilde I(\br,\sigma)$ is responsible for the sign of $\delta n$ dependence on $\br$ direction and, in turn, depends on $|\br|$ through a power law.

\begin{figure}
\includegraphics[width=.95\columnwidth]{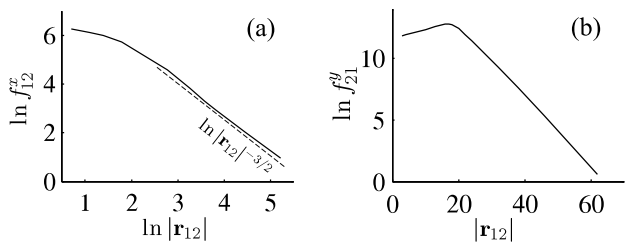}
\caption{\label{fig:dist}
Asymptotic behavior of dissipative forces (a) $f_{12}^x$ (longitudinal alignment) and (b) $f_{12}^y$ (transverse alignment) at large inter-obstacle separation $r_{12}$. The slope on (a) corresponds to the asymptotics $f_{12}^x\sim r_{12}^{-3/2}$. Equilibrium concentration $n_0=0.8$, external field $\bg$ ($|\bg|=0.5$) is directed along the $x$-axis, the impermeable circular obstacles are of radius $a=7$ (in units of $\ell$), forces are in units of $kT/\ell$.
}
\end{figure}

Behavior of $\tilde I(\br, \bq)$ in the case of a spherical obstacle is defined by the asymptotics of the Bessel function $K_{m+\frac{1}{2}}(qr)$ \cite{abramowitz_handbook_1965}:
\begin{equation}\label{eq:A2:I}
  \delta n^0\approx\sqrt{2\pi}a^2\frac{e^{-qr(1-\beta\cos\theta)}}{r}\tilde I(\br,\bq),
\end{equation}
\begin{equation}\label{eq:A2:Is}
I\approx\sum_{m=0}\alpha_m\left(1+\frac{m^2+m}{2qr}+
\cdots\right)\frac{I_{m+\frac{1}{2}}(qa)}{\sqrt{qa}}P_m(\cos\theta),
\end{equation}
where $\alpha_m=\alpha_m(qa)$ depends only on the obstacle radius $a_k$ and external field $\bq$. The coefficients $\alpha_m$ are from the Legendre polynomials expansion $\mu^0(\theta)=e^{\beta q a \cos\theta}\sum_{n=0}^\infty \alpha_n P_n (\cos\theta)$ at the obstacle surface and can be obtained as a solution of equation (\ref{eq:A2:intGeen}), $\theta$ is the angle between $\br$ and $\bg$ and $\beta=(1/2-n_0)/|1/2-n_0|=\pm1$. Distribution $\delta n^0(\br)$ for an isolated circular inclusion in 2D case was obtained in \cite{kliushnychenko_blockade_2014}. In the particular case of $n_0<1/2$, the dipole approximation gives the following distributions for the gas perturbation:
ahead of the obstacle, $\bq\br=-qr$,
\begin{equation}
\delta n(r)\approx be^{-2qr} \left\{ \left[3c+1\right](qr)^{-\frac{1}{2}}+\frac{3}{8}\left[ c+1\right](qr)^{-\frac{3}{2}}\right\},
\end{equation}
and behind it, $\bq\br=qr$,
\begin{equation}
\delta n(r)\approx -b \left\{ (qr)^{-\frac{1}{2}}+\frac{3}{8}\left[ 2c+1 \right](qr)^{-\frac{3}{2}}\right\}.
\end{equation}
Here, constants $c=2\left[ 3(qa)K_1(qa)I_2(qa)\right]^{-1}$ and $b=\sqrt{8\pi}n_0(1-n_0)|1-2n_0|^{-1}I_2(qa)K^{-1}_2(qa)$  are expressed in terms of the modified Bessel functions $I_n$ and $K_n$; $a$ is the radius of the impermeable obstacle ($\lambda\rightarrow1$). In the  case of the point-like inclusion, $qa\sim q\ell\ll 1$, this method gives $\delta n \sim e^{-2qr}r^{-1/2}$ for a region ahead of the inclusion and $\delta n \sim -r^{-3/2}$ for the tail asymptotics. This is in qualitative agreement with the numerical results \cite{kliushnychenko_blockade_2014} and coincides with the asymptotic behavior of the wake relaxation for a moving intruder \cite{benichou_stokes_2000}.
The general form of the dissipative force in 2D case is analogous to Eq.~(\ref{eq:A2:f2}):
\begin{equation}
  \bff_{ki}\propto-\frac{e^{-q|\br_{ki}|+\bq\br}}{|\br|^{1/2}}\int_{S_k}\bn(\bx_k)I(\br_{ki},\bq,\bx_k)\,\id\bx_k.
\end{equation}
It is easy to show that for longitudinal alignment $\bu_{12}(\br_{12}) = -\br_{12}/|\br_{12}|^2$ and the leading asymptotic behavior $f_{12}^x\approx A |\br_{ki}|^{-3/2}$, that is in agreement with numerical result for nonlinear Eq.~(\ref{balance+}), see Fig.~\ref{fig:dist}(a),
when the distance between obstacles $|\br_{ki}|$ is much larger than their radii $a_i$. The form-factor $A$ depends only on external field $\bg$ and the obstacle radius $a$. For the transverse alignment, the force leading asymptotics behaves exponentially, $\ln f^y_{12}\propto-q|\br_{12}|+\cdots$, that is also in qualitative agreement with numerical result, see Fig.~\ref{fig:dist}(b).


\begin{thebibliography}{99}
\bibitem{comment}
In particular, it gives symmetrical form of probe-particle wake (with the same damping of medium perturbation ahead of the particle and behind it), while the typical form of wake profile occurring for a moving inclusion is asymmetrical (with a localized dense region ahead of inclusion and an extended depletion region behind it).
\bibitem{a}
While defining the average $\overline {(\dots)_j}$ we omitted the normalization constant. Also, gas density distribution $n(\br)$ can be described by a perturbation $\delta n$ of a non-uniform equilibrium distribution of the gas with inclusions $n(\br)\approx n_0[1-U(\br)]+\delta n(\br)$, not the equilibrium concentration of the inclusion-free gas $n_0$, Eq. (\ref{eq:A1:01.5}). Since equations for $\delta n(\br)$ and $\Omega (\br)$ turn to be similar to (\ref{eq:A1:07}) and (\ref{eq:A1:08}), our qualitative results are retained.
\end{thebibliography}
\end{document}